\documentclass[twocolumn, pra,showpacs,superscriptaddress]{revtex4}
\usepackage{graphicx}
\usepackage{dcolumn}
\usepackage{bm}
\usepackage{amsmath}

\begin{document}

\title{The weakening of fermionization of one dimensional spinor Bose gases induced by spin-exchange interaction}
\author{Yajiang Hao}
\email{haoyj@ustb.edu.cn}
\affiliation{Department of Physics, University of Science and Technology Beijing, Beijing
100083, China}
\date{\today }

\begin{abstract}
We investigate the ground state density distributions of anti-ferromagnetic spin-1 Bose gases in one dimensional harmonic potential in the full interacting regimes. The ground state is obtained by diagonalizing the Hamiltonian in the Hilbert space composed of the lowest eigenstates of noninteracting Bose gas and spin components. The study reveals that in the situation of weak spin-dependent interaction the total density profiles evolve from Gaussian-like distribution to a  Fermi-like shell structure of $N$  peaks with the increasing of spin-independent interaction. While the increasing spin-exchange interaction always weaken the fermionization of density distribution such that the total density profiles show shell structure of less peaks and even show single peak structure in the limit of strong spin-exchange interaction. The weakening of fermionization results from the formation of composite atoms induced by spin-exchange interaction. It is also shown that phase separation occurs for the spinor Bose gas with weak spin-exchange interaction, meanwhile strong spin-independent interaction.
\end{abstract}

\maketitle

%\author{Yunbo Zhang}
%\affiliation{Department of Physics and Institute of Theoretical
%Physics, Shanxi University, Taiyuan 030006, P. R. China}
%\author{Shu Chen}
%\affiliation{Beijing National
%Laboratory for Condensed Matter Physics, Institute of Physics,
%Chinese Academy of Sciences, Beijing 100080, P. R. China}

%05.30.Jp Boson systems (for static and dynamic properties of Bose-Einstein condensates,
%         see 03.75.Hh and 03.75.Kk; see also 67.10.Ba Boson degeneracy in quantum fluids)¡£
%03.75.Hh Static properties of condensates; thermodynamical, statistical, and structural properties
%03.75.Mn Multicomponent condensates; spinor condensates¡£
%67.85.-d Ultracold gases, trapped gases

\section{Introduction}
When the  Bose-Einstein condensate (BEC) was confined in an optical trap rather than in a magnetic trap, the atomic spin freedom degrees were liberated and the spinor BEC \cite{T.L.Ho,T.L.Ho2,Jap} was realized. In contrast to the single component BEC, the spinor BEC manifest rich physics for its rich spin textures \cite{PhysRep}.  In contrast to the two-species BEC, the spinor BEC has special internal spin-mixing dynamics \cite{Law}, in which two atoms of $m=0$  can coherently scatter into one atom of $m=1$ and one atom of $m=-1$, or vice versa. The realization of spinor BEC\cite{Ketterle98PRL,Champman2001PRL} has offered us a populated platform to investigated the effects related with the spin in the conventional condensed matter physics. With the spinor BEC many experimentalists and theorists have investigate the formation of spin domain \cite{spindomain,LZhou}, the exploration of novel quantum phase \cite{Sadler,QGuPRL}, the magnetism \cite{QGu,QGuPRB,AVinit,Higbie}, spin textures \cite{PhysRep,SYi}, the spinor vortex \cite{Vortex,Villasenor,Seo,Manni}, the effect of spin-orbit coupling \cite{HZhai, Lan}, and spin-mixing dynamics \cite{Law,Huang,He,WX Zhang}, etc \cite{RMP}. Besides the spinor BEC of $F=1$, the spinor BEC of high spin were also paid many attentions to \cite{Qi,Mithun,Gautam}.

With the development of the experiment technique, the ultracold atom gas can be confined in highly anisotropic trap and optical lattices such that the strong correlated quantum gas in low dimension become available \cite{Paredes,Toshiya,Ketterle,Single1D}.  The realization of low dimensional quantum gas fascinated many researchers because  before this the presently real physical systems are just "toy models" in textbooks, and now we can understand important physics by parameter-free comparison of theoretical prediction with measurement. In experiment we can realize not only the ground state of one dimensional (1D) Bose gas, but also the excited state \cite{STG}. The rapid experiment progresses stimulated significant interest in the low dimensional quantum gas for its strongly correlated effect  \cite{RMP2011,RMP2012,RMP2013}.

With the Feschbach resonance and confined induced resonance technique \cite{FR,CIR} the inter-atomic interaction strength can be tuned in the full interacting regime from infinite attraction to infinite repulsion. In the weakly interacting regime, the properties of 1D quantum gas can be described in the mean-field theory. The reduced 1D Gross-Pitaevskii equation can be utilized to investigate its ground state properties. For the 1D quantum gas in strongly interacting regime, the strong quantum fluctuations become remarkable and we have to turn to the non-perturbed method beyond mean-field theory, for example, the Bethe ansatz \cite{Lieb}, exact numerical diagonalization method \cite{Deuretzbacher,Hao2009}, Bose-Fermi mapping method \cite{Girardeau2}, multi-configuration mean-field theory \cite{MCTDHB}, density-matrix renormalization group \cite{DMRG}, etc. It has been shown that the weakly interacting Bose gas exhibits the Gaussian-like Bose distribution, while the strongly interacting Bose gas, i.e., Tonks-Girardeau gas, is fermionized and exhibit the same density distribution as the spin-polarized fermions \cite{Hao2006}. For the strongly interacting multi-component quantum mixture, composite fermionization are exhibited \cite{Hao2009,Zollner2008,HaoCPL,Fang}. In addition it was shown that the quantum mixtures with internal freedom degree display the "spin-charge" separation \cite{Fush,Kleine} and magnetic order \cite{GuanXW,Dehkharghani}, etc.

So far the investigation on 1D quantum gas mixture mainly focus on two-component of Bose gas, Bose-Fermi mixture  and Fermi gas of internal freedom degree, while the spin-1 spinor Bose gas was not systematically studied. F. Deuretzbacher etc. presented the exact solution of spin-1 1D Bose gas in the infinite strong repulsion limit with the Bose-Fermi mapping method \cite{DeuretzbacherPRL,HPu}. The ground state of 1D spinor Bose gas in the full repulsive interaction regime were also obtained \cite{Wang,HaoSpinor}, but the previous results were only restricted on the case with weak spin-dependent (spin-exchange) interaction \cite{HaoSpinor} and the case of equal spin-dependent interaction and spin-independent interaction \cite{Wang}. The effect of strong spin-exchange interaction and the competition between spin-dependent interaction and spin-independent interaction on rich physics of 1D spinor Bose gas are subjects to study.

The present paper mainly focus on the effect of spin-(in)dependent interactions on the ground state density distributions, both of which can be tuned from weak interaction to strong interaction. In this situation the spin-(in)dependent interaction strength might be arbitrary value and no exact solution can be employed even in 1D.  We will extend the previous developed diagonalization method \cite{Deuretzbacher,Hao2009,HaoEPJD} to the investigation on the ground state of 1D spinor Bose gas. By diagonalizing the Hamiltonian in the Hilbert space composed of energetically lowest eigenstates of the non-interaction Bose gas and spin states, the ground state wavefunction and therefore the density distribution of each components will be obtained. Compared with the single component Bose gas and two-component mixtures, the internal freedom degree of spinor Bose gas results in a quite large Hilbert space. To reduce the dimension of Hilbert space to a feasible size for numerical diagonalization, we execute the evaluation in the subspace of special total spin.

The paper is organized as follows. In Sec. II, we briefly review the 1D spinor model in a harmonic trap and its second-quantized Hamiltonian in the Hilbert space. The numerical diagonalization method will also be introduced. In Sec. III, we present results of the ground-state density distributions for the 1D spinor gases in the full repulsive interaction regime. A brief summary is given in Sec. V.

\section{The model and method}
We consider $N$ identical bosons with mass $m$ and spin 1 confined in an external potential $V(x)=\frac12m\omega^2x^2$. In the binary collision the total spin is conserved and it is 0 or 2 for two interacting atoms of spin 1. The 1D effective interaction constant $g_f$ is related to the $s$-wave scattering length of the total spin-$f$ channel $a_f$ ($f$=0, 2) \cite{Olshanii,Dunjko,Petrov,HaoSpinor}
\begin{equation*}
g_f=\frac{4\hbar^2a_f}{ma_{\bot}^2}(1-\mathcal{C}\frac{a_f}{a_{\bot}})^{-1},
\end{equation*}
where $\mathcal{C}=1.4603$, and $a_{\bot}=\sqrt{\hbar/m\omega_{\bot}}$ with transverse trapping frequency $\omega_{\bot}$. With Feshbach resonance technique and confined induced resonance we can tune the effective one-dimensional interaction in the full interacting regime.

For the spinor gas its Hamiltonian in the second quantized form can be formulated as \cite{T.L.Ho,T.L.Ho2,Jap,Law}
\begin{eqnarray}
\hat{\mathcal{H}} &=&\int dx\left[ \hat{\Psi}_{\alpha }^{\dag }(x)(-\frac{
\hbar ^{2}}{2m}\frac{d^{2}}{dx^{2}}+V(x))\hat{\Psi}_{\alpha }(x)\right.  \nonumber \\
&&\left. +\frac{c_{0}}{2}\hat{\Psi}_{\alpha }^{\dag }(x)\hat{\Psi}_{\beta
}^{\dag }(x)\hat{\Psi}_{\beta }(x)\hat{\Psi}_{\alpha }(x)\right.  \\
&&\left. +\frac{c_{2}}{2}\hat{\Psi}_{\alpha }^{\dag }(x)\hat{\Psi}_{\alpha
\prime }^{\dag }(x)\mathbf{F}_{\alpha \beta }\cdot \mathbf{F}_{\alpha \prime
\beta \prime }\hat{\Psi}_{\beta \prime }(x)\hat{\Psi}_{\beta }(x)\right].\nonumber
\end{eqnarray}
Here the interaction is composed of spin-independent part and spin-dependent part, which are related with the interaction constants $c_0=\frac{g_0+2g_2}3$ and $c_2=\frac{g_2-g_0}3$, respectively. $\mathbf{F}$ is the spin-1 matrix. The field operator $\hat{\Psi}_{\alpha}^{\dag}(x)$ [$\hat{\Psi}_{\alpha}(x)$] creates (annihilates) an $\alpha$-component atom at the position $x$.

The operator $\hat{\Psi}_{\alpha}(x)$ can be expanded in the Hilbert space, which is constructed by the single particle wavefunctions (orbital) of a particle in harmonic trap $\phi _i(x)=\frac {H_i(x)\exp (-x^{2}/2)}{\pi ^{1/4}\sqrt{2^{i }i !}}$ with Hermite polynomial $H_i(x)=i !\sum_{k=0}^{[i /2]}(-1)^{k}(2x)^{i -2k}/k!/(i -2k)!$, i.e.,
\begin{equation*}
\hat{\Psi}_{\alpha }(x)=\sum_{i=1}^L\phi _{i}\left( x\right) \hat{b}_{i\alpha }
\end{equation*}
with $\hat{b}_{i\alpha }$ being the destruction operator of the $\alpha$-component atom in the $i$th orbital. The Hilbert space is complete as long as the orbital number $L$ is large enough. Thus the many body Hamiltonian will be expressed as
\begin{eqnarray*}
H &=&\sum_{i,\alpha }\mu _{i}\hat{b}_{i\alpha }^{\dagger }\hat{b}_{i\alpha }+%
\frac{c_{0}}{2}\sum_{\alpha \beta }\sum_{ijkl}I_{ijkl}\hat{b}_{i\alpha
}^{\dagger }\hat{b}_{j\beta }^{\dagger }\hat{b}_{k\beta }\hat{b}_{l\alpha }
\\
&&+\frac{c_{2}}{2}\sum_{\alpha \beta ;\alpha ^{\prime }\beta ^{\prime
}}\sum_{ijkl}I_{ijkl}\hat{b}_{i\alpha }^{\dagger }\hat{b}_{j\alpha ^{\prime
}}^{\dagger }\left( \mathbf{F}\right) _{\alpha \beta }\cdot \left( \mathbf{F}%
\right) _{\alpha ^{\prime }\beta ^{\prime }}\hat{b}_{k\beta ^{\prime }}\hat{b%
}_{l\beta },
\end{eqnarray*}
where $\mu _{i}=\left( i+\frac{1}{2}\right) \hbar \omega $ and $I_{ijkl} =\int dx\phi _{i}\left( x\right) \phi _{j}\left( x\right)\phi _{k}\left( x\right) \phi _{l}\left( x\right)$.

By diagonalizing the Hamiltonian in the Hilbert space we can obtain the ground state wavefunction and therefore the interested quantities. For the spinor bose gas the Hilbert space is composed of orbital wavefunctions and spin component. The dimension of the Hilbert space for spin-1 bose gas is $C_{N+3L-1}^N$. As the atomic interaction is strong, more orbitals should be considered and $L$ should be large enough. In this situation the dimension of Hilbert space will be so large that the numerical diagonalization is impossible to be executed. In our evaluation we will diagonalize the Hamiltonian in the subspace of total spin being conserved. The dimension of Hilbert space can be reduced enormously in this way. In the preset paper the space of magnetization $M=0$ will be investigated. The dimension of Hilbert space spanned by 20 orbitals is 136955 for $N=4$, which is feasible for numerical diagonalization.

\section{The density distribution of ground state}
In the following we will study the density distributions of ground state in the full interacting regime for anti-ferromagnetic spinor gas ($c_2>0$). The density distribution of $\alpha$-component can be evaluated by
\begin{eqnarray*}
n _{\alpha }\left( x\right)  &=&\left\langle GS\left\vert \hat{\Psi}%
_{\alpha }^{\dag }(x)\hat{\Psi}_{\alpha }(x)\right\vert GS\right\rangle  \\
&=&\sum_{ij}\phi _{i}\left( x\right) \phi _{j}\left( x\right) \left\langle
GS\left\vert \hat{b}_{i\alpha }^{\dagger }\hat{b}_{j\alpha }\right\vert
GS\right\rangle.
\end{eqnarray*}
The total density profile is given by $n(x)=\sum_{\alpha}n_{\alpha}(x)$. For simplicity in the following presentation we will use the dimensionless interaction $U_i$ ($i=0,2$ and $U_i=c_i/a_0$ with the harmonic length $a_0=\sqrt{\hbar/m\omega}$) to denote the spin-(in)dependent interaction strength.

\subsection{Fermionization of anti-ferromagnetic spinor bose gases}

\begin{figure}[tbp]
\includegraphics[width=3.0in]{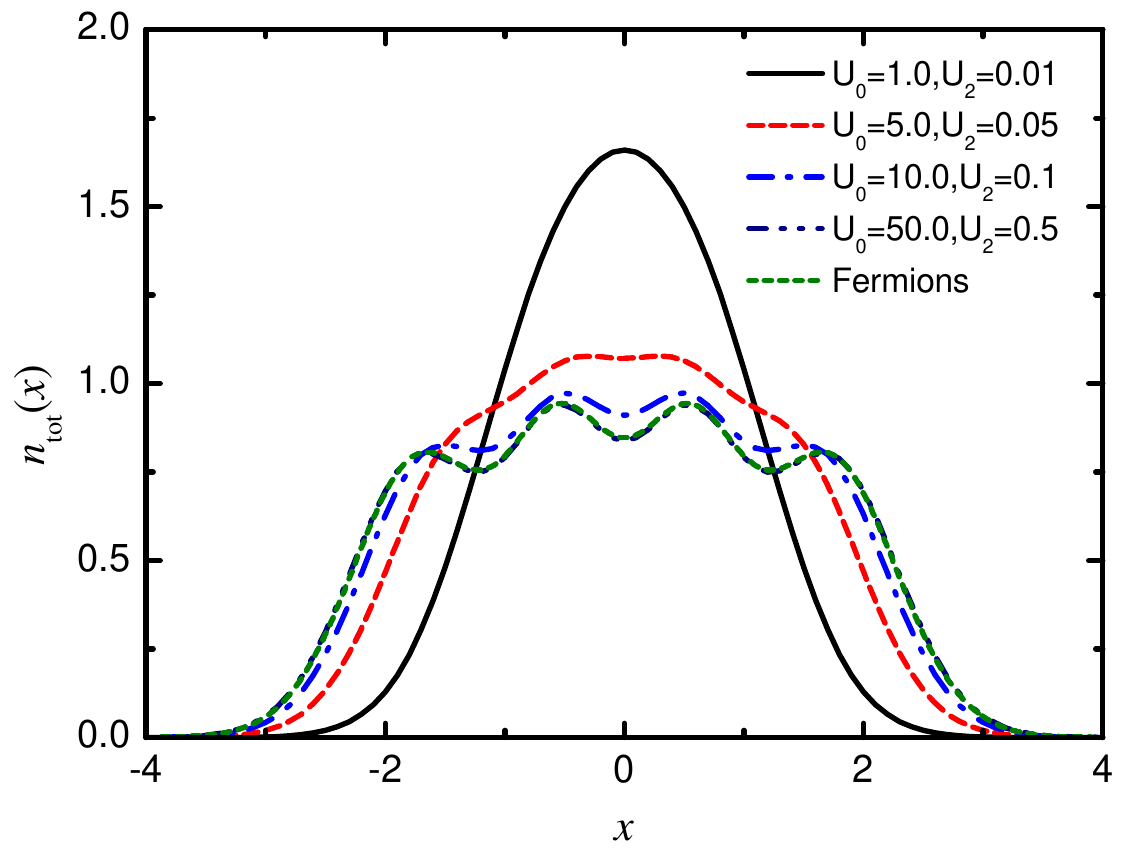}
\caption{(color online) Total density distribution of the ground state for $N=4$ with $U_0=100U_2$. The density profiles of spin-polarized fermions is plotted in short dash lines.}
\end{figure}

Usually the spin-dependent interaction $U_2$ is two order of magnitude lower than the spin-independent interaction $U_0$. In Fig. 1 we display the total density profiles of ground state for 1D spinor bose gases with $U_0=100U_2$. In the weak interaction regime ($U_0=1$ here) the density profiles exhibit the typical Bose distribution of single peak, in which atoms populate in the center of the trap in large probability. With the increase of interaction Bose atoms prefer to seperate each other and the regime away from the potential center will be occupied in the increased probability. The half-width of density profiles become larger. In the strong interaction regime ($U_0 \ge 10$) the density distribution display the shell structure of $N$ peaks. The spinor bose gas is fermionized and exhibit the same density distribution as that of TG gas of $N$ Bosons. Both of them are identical to the density distributions of $N$ spin-polarized Fermions.

\begin{figure}[tbp]
\includegraphics[width=3.0in]{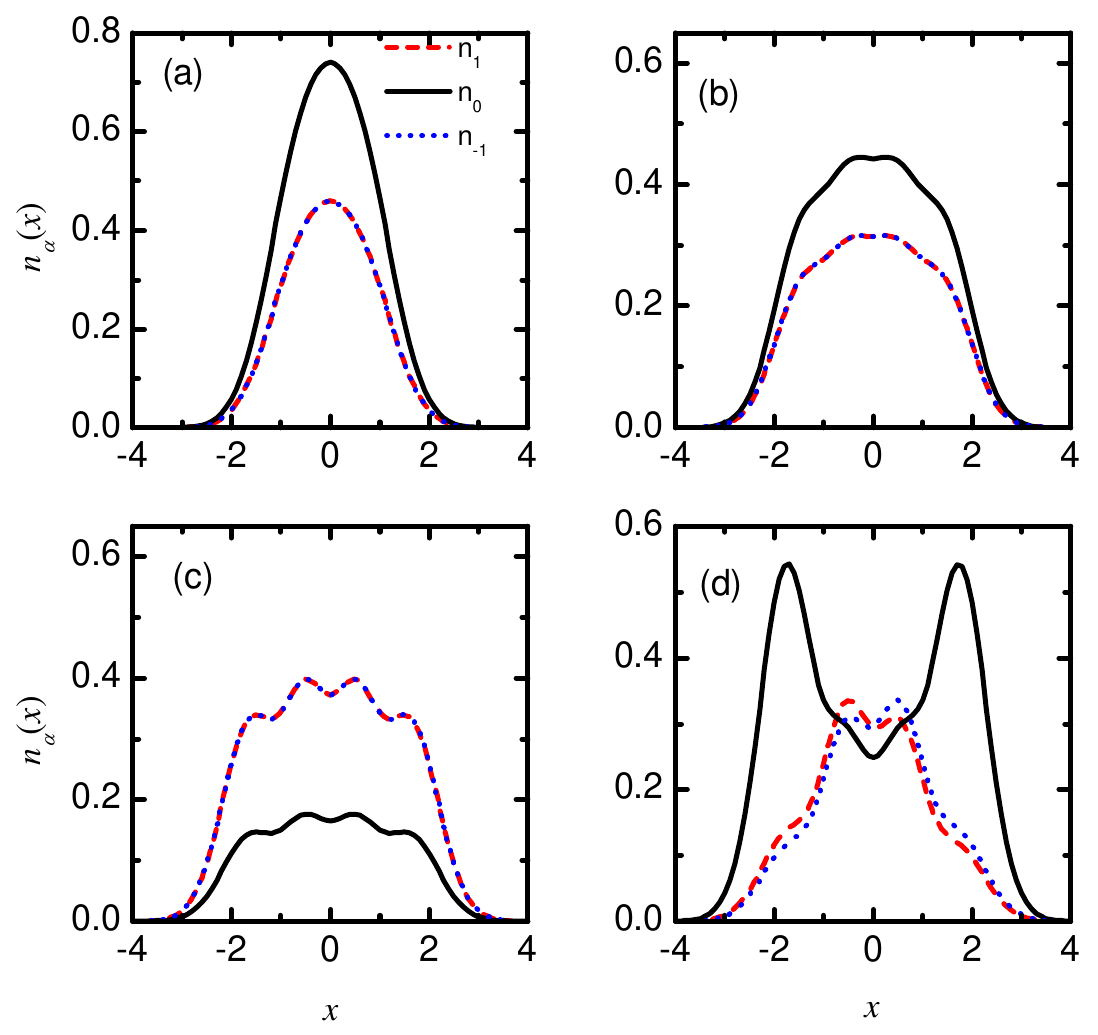}
\caption{(color online) Density distributions of each component of the ground state for $N=4$ with $U_0=100U_2$. (a) $U_0$=1.0 and $U_2$=0.01; (b) $U_0$=5.0 and $U_2$=0.05; (c) $U_0$=10.0 and $U_2$=0.1; (d) $U_0$=50.0 and $U_2$=0.5.}
\end{figure}

The corresponding density profiles of each component are plotted in Fig. 2. As shown in Fig. 2a-2c, in the weak interaction regime and middle interaction regime, three components overlap each other, and spin-1 and spin-(-1) components have the identical profiles. As the interaction increases, each component exhibits the same fermionized behaviour. In the TG limit (Fig. 2d), three components behave in different way and the phase separation appears. The 0-component display the structure of double peak at the regime away from the potential center and its density profiles is symmetry about the center of the trap. The $\pm1$-component prefer to distribute in the center regime but they do not completely overlap each other and their respective profiles are not symmetric about the center of the trap. The peak of one component appears on the left side of the trap, and that of the other component in the right side of the trap.

\subsection{The weakening of fermionization}
\begin{figure}[tbp]
\includegraphics[width=3.0in]{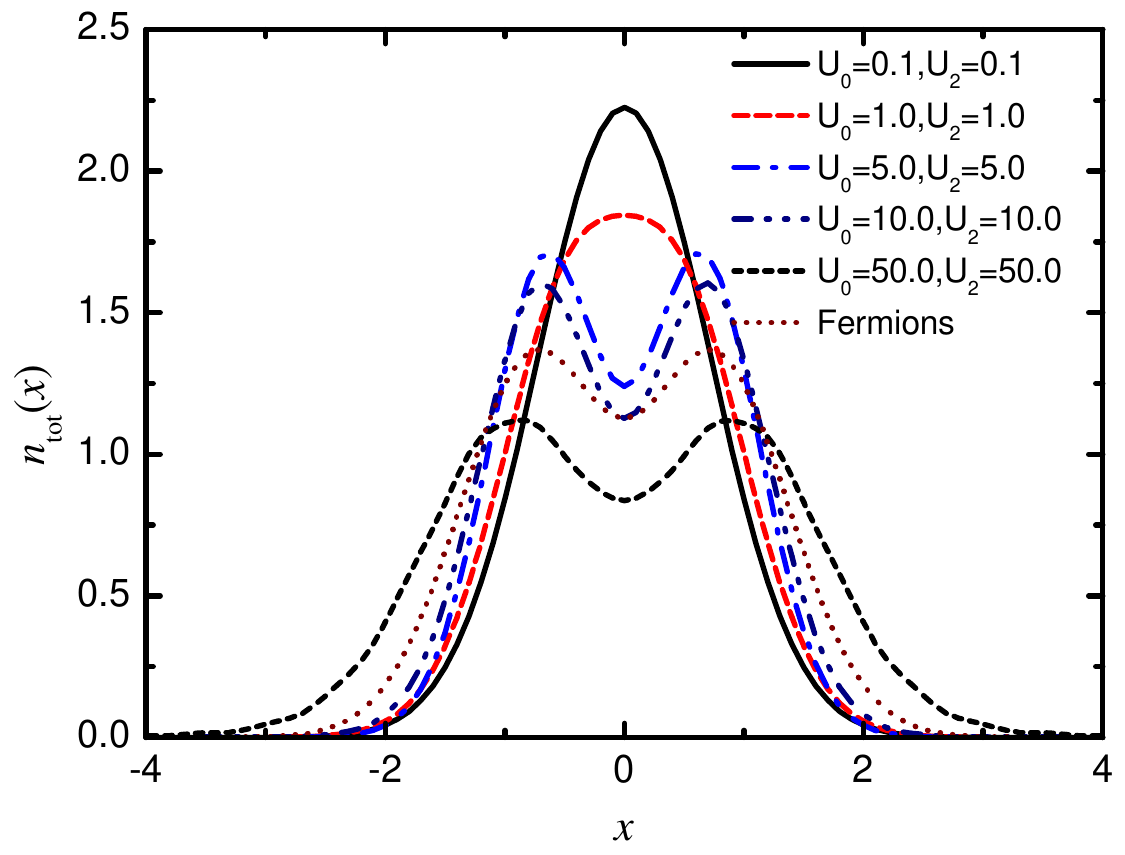}
\caption{(color online) Total density distribution of the ground state for $N=4$ with $U_0=U_2$.}
\end{figure}

It is interesting to investigate the effect of strong spin-exchange interaction in the spinor Bose gases. That is to say, the spin-dependent interaction is comparable to or even stronger than the spin-independent interaction. In Fig. 3 we display the total density profiles of three components when the spin-dependent interacting strength is equal to spin-independent interacting strength ($U_0=U_2$). It is shown that with the increase of interaction the single peak structure of the density distribution embodying the Boson property disappear and the multi-peaks structure embodying the Fermion property appear. The stronger interaction leads to the atomic distribution in wider regimes. Compared with the case of weak spin-exchange interaction two peaks have been shown in the middle interaction regime ($U_0=5.0$) rather than four peaks shown only in the TG limit. Another difference is that only two peaks are exhibited even in the strong interaction limit. The spinor Bose gas do not exhibit the complete fermionization behaviour same as the situation of weak spin-exchange interaction. The fermionization behaviour resulting from the strong spin-independent interaction become weak because of the strong spin-dependent interaction. It can be interpreted as the fermionization of two composite atoms, which are the result of atom pairing for the strong spin-exchange interaction. The composite atoms behave differently from the single atom. As a comparison, the density profile of two spin-polarized fermions is plotted in Fig. 3, which is different from the distribution of two composite atoms quantitatively. The density distribution has been normalized to the particle number for comparison.

\begin{figure}[tbp]
\includegraphics[width=3.0in]{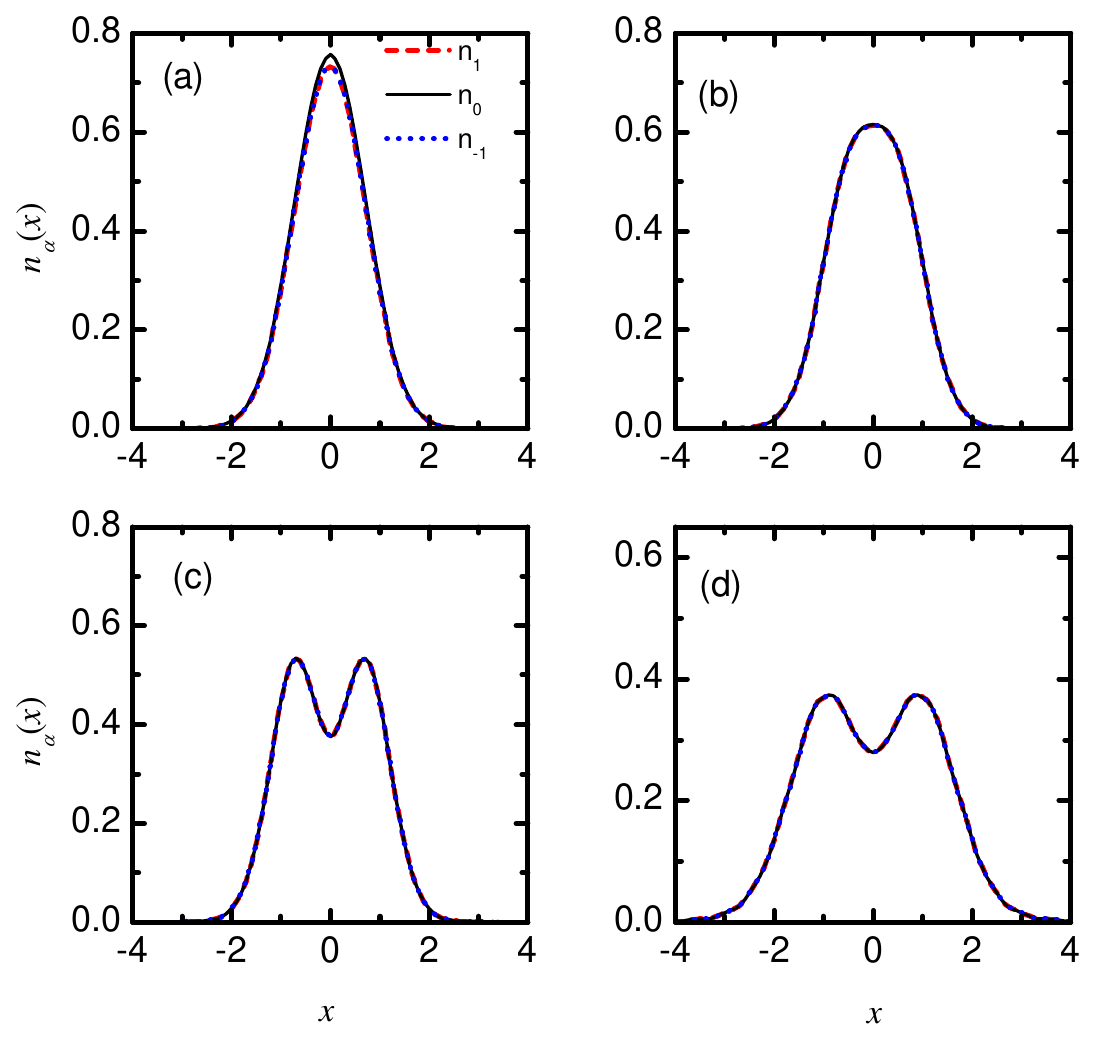}
\caption{(color online) Density distribution of each component of the ground state for $N=4$ with $U_0=U_2$. (a) $U_0=U_2$=0.1; (b) $U_0=U_2$=1.0; (c) $U_0=U_2$=10.0; (d) $U_0=U_2$=50.0.}
\end{figure}

The corresponding density distributions of each component are displayed in Fig. 4. It is shown that in the full interacting regime three components overlap each other. They have exactly same behaviour except that the density profile of 0-component is different from those of $\pm$1-component in weak interacting regime ($U_0=U_2=1.0$).

\subsection{The breakdown of fermionization of spinor Bose gases}
\begin{figure}[tbp]
\includegraphics[width=3.0in]{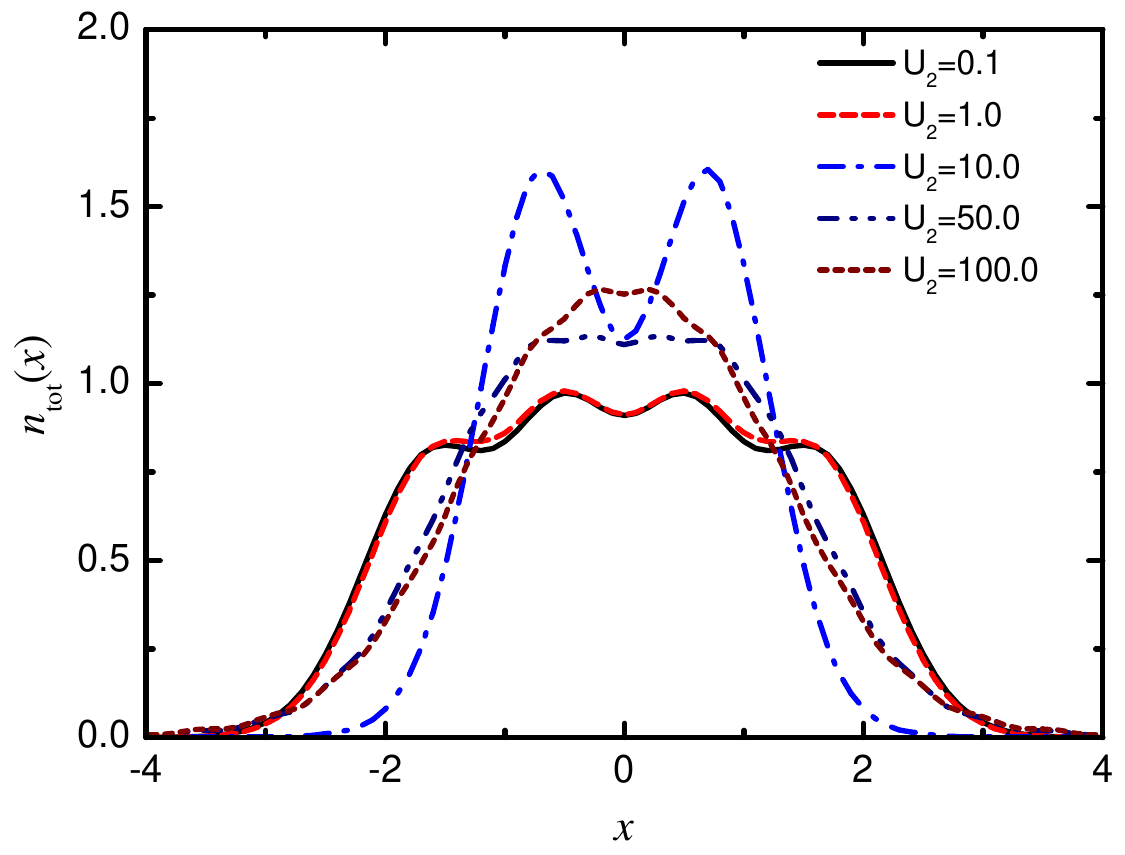}
\caption{(color online) Total density distribution of the ground state for $N=4$ with $U_0$=10.0.}
\end{figure}

As shown in the previous section, the effect of strong spin-exchange interaction is to weaken the fermionization of Bose gases. In order to systematically investigate the weakening effect, in this subsection, we fix the spin-independent interaction $U_0=10.0$ and tune the spin-dependent interaction $U_2$ from 0.1 to 100. The total density profiles are shown in Fig. 5 for different spin-exchange interaction. For weak spin-exchange interaction the density profiles show typical shell structure of TG gases with $4$ peaks (for example, $U_2$=0.1 and 1.0 here). As the spin-dependent interaction is comparable with the spin-independent interaction ($U_0=U_2$) the shell structure of 4 peaks change into the shell structure of 2 peaks for the formation of composite particle of two atoms. With the further increase of spin-exchange interaction ($U_2=100$), the shell structure of 2 peaks disappear and there is only one peak in the density distribution. Thus the fermionization is broken down completely.

\begin{figure}[tbp]
\includegraphics[width=3.0in]{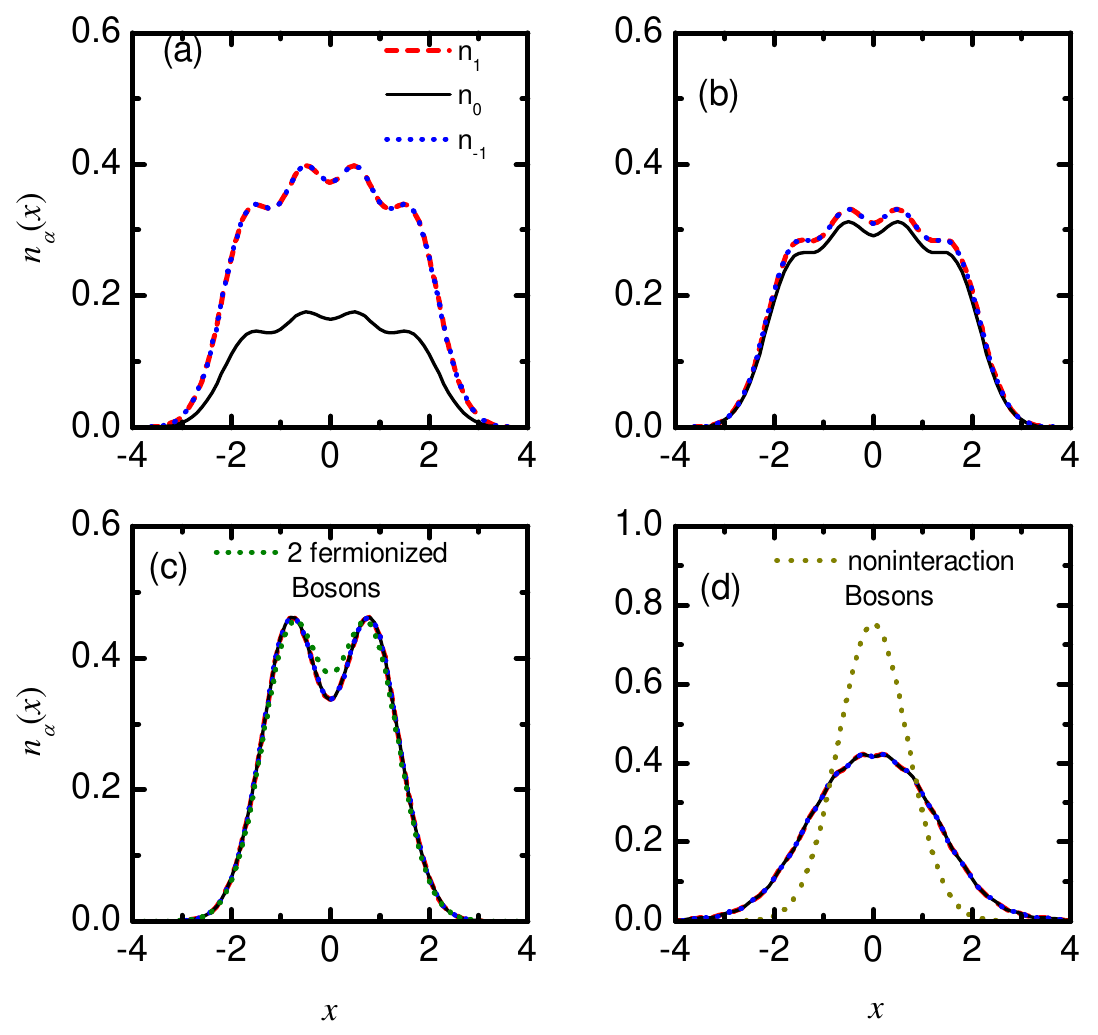}
\caption{(color online) Density distribution of each component of the ground state for $N=4$ with $U_0=10.0$. (a) $U_2$=0.1; (b) $U_2$=1.0; (c) $U_2$=5.0; (d) $U_2$=100.0. The density profiles of two fermionized Bose atoms and noninteracting Bose gases are plotted in short dash line and short dot line, respectively.}
\end{figure}

The density profiles of each component for $U_0=10.0$ are plotted in Fig. 6. In the full interacting regime each components overlap each other. With the increase of spin-exchange interaction each component change from 4 peaks structure into 2 peaks structure, and further into single peak structure. In the weak spin-exchange interaction ($U_2$=0.1 and 1.0), atomic number in the spin-1 component is same as that in the (-1)-component, while the the atomic number in 0-component is less than that in $\pm1$-component. As the spin-dependent interaction become strong, three components have the same atom number. As comparisons, we plot the shell structure of two peaks for two fermionized Bose atoms in Fig. 6c and the single peak distribution of non-interacting Bose atoms in Fig. 6d, for both of which the particle number is normalized to $N/3$ for comparison. It is shown that the double peaks structure of spinor bose gas is similar to but not same as the fully fermionized single component Bose gases, and the single peak structure of spinor bose gas in the strong spin-exchange interaction is also different from the non-interacting Bose gas.

%\begin{figure}[tbp]
%\includegraphics[width=3.0in]{N4density_totalU010Mz5.eps}
%\caption{(color online) Total density distribution of the ground state for $N=4$ with $U_0$=10.0 and $M=0.5$.}
%\end{figure}
%
%\begin{figure}[tbp]
%\includegraphics[width=3.0in]{N4density_componentU010Mz5.eps}
%\caption{(color online) Density distribution of each component of the ground state for $N=4$ with $U_0=10.0$ and $M$=0.5. (a) $U_2$=0.1; (b) $U_2$=1.0; (c) $U_2$=10.0; (d) $U_2$=50.0.}
%\end{figure}

\section{Summary}

In the present paper we investigate the density distributions of anti-ferromagnetic spin-1 Bose gas in a 1D harmonic trap in the full interacting regime from the weak interaction to strong interaction for both spin-independent interaction and spin-dependent interaction. By diagonalizing the Hamiltonian in the Hilbert space composed of the lowest eigenstates of single particle and spin components, we obtain the ground state and its density profiles.

For weak spin-dependent interaction, the total density profiles of spinor bose gas transit from Gaussian-like Bose distribution to Fermi-like distribution with the increase of spin-independent interaction. Each component show the same fermionization transition, but in the strong spin-indepdent interacting limit three components prefer to separate and phase separation appear. As the spin-exchange interaction is comparable to the spin-independent interaction, the increasing spin-independent interaction also induce the fermionized density profiles of shell structure, but only $N/2$ peaks appear rather than the $N$-peak structure in the case of weak spin-exchange interaction. If the spin-exchange interaction increase further and is even stronger than spin-independent interaction, the shell structure embodying fermionization disappear and single-peak density profiles are shown. The disappearance of $N$-peaks shell structure of the density profiles can be interpreted as the result from the formation of composite atom induced by the strong spin-exchange interaction. The comparison with the density distribution of spin-polarized fermions shows that the behaviors of the composite atoms are same as the spin-polarized fermions qualitatively rather than quantitatively.   Another effect of strong spin-exchange interaction is that each component exhibits the exactly same behaviors.

\begin{acknowledgments}
The authors acknowledge the NSF of China (Grant No. 11004007) and ``the Fundamental Research Funds for the Central Universities".
\end{acknowledgments}

%$\dagger$ Electronic address: schen@aphy.iphy.ac.cn

\end{document}